\documentstyle[twoside,epsfig]{article}
\catcode`\@=11
\long\def\@makefntext#1{
\protect\noindent \hbox to 3.2pt {\hskip-.9pt
$^{{\eightrm\@thefnmark}}$\hfil}#1\hfill}		%CAN BE USED

\def\thefootnote{\fnsymbol{footnote}}
\def\@makefnmark{\hbox to 0pt{$^{\@thefnmark}$\hss}}	%ORIGINAL

\def\ps@myheadings{\let\@mkboth\@gobbletwo
\def\@oddhead{\hbox{}
\rightmark\hfil\eightrm\thepage}
\def\@oddfoot{}\def\@evenhead{\eightrm\thepage\hfil
\leftmark\hbox{}}\def\@evenfoot{}
\def\sectionmark##1{}\def\subsectionmark##1{}}

\oddsidemargin=\evensidemargin
\renewcommand{\thefootnote}{\fnsymbol{footnote}}

\newcounter{sectionc}\newcounter{subsectionc}\newcounter{subsubsectionc}
\renewcommand{\section}[1] {\vspace{12pt}\addtocounter{sectionc}{1}
\setcounter{subsectionc}{0}\setcounter{subsubsectionc}{0}\noindent
	{\tenbf\thesectionc. #1}\par\vspace{5pt}}
\renewcommand{\subsection}[1] {\vspace{12pt}\addtocounter{subsectionc}{1}
	\setcounter{subsubsectionc}{0}\noindent
	{\bf\thesectionc.\thesubsectionc. {\kern1pt \bfit #1}}\par\vspace{5pt}}
\renewcommand{\subsubsection}[1] {\vspace{12pt}\addtocounter{subsubsectionc}{1}
	\noindent{\tenrm\thesectionc.\thesubsectionc.\thesubsubsectionc.
	{\kern1pt \tenit #1}}\par\vspace{5pt}}
\newcommand{\nonumsection}[1] {\vspace{12pt}\noindent{\tenbf #1}
	\par\vspace{5pt}}

\newcounter{appendixc}
\newcounter{subappendixc}[appendixc]
\newcounter{subsubappendixc}[subappendixc]
\renewcommand{\thesubappendixc}{\Alph{appendixc}.\arabic{subappendixc}}
\renewcommand{\thesubsubappendixc}
	{\Alph{appendixc}.\arabic{subappendixc}.\arabic{subsubappendixc}}

\renewcommand{\appendix}[1] {\vspace{12pt}
        \refstepcounter{appendixc}
        \setcounter{figure}{0}
        \setcounter{table}{0}
        \setcounter{lemma}{0}
        \setcounter{theorem}{0}
        \setcounter{corollary}{0}
        \setcounter{definition}{0}
        \setcounter{equation}{0}
        \renewcommand{\thefigure}{\Alph{appendixc}.\arabic{figure}}
        \renewcommand{\thetable}{\Alph{appendixc}.\arabic{table}}
        \renewcommand{\theappendixc}{\Alph{appendixc}}
        \renewcommand{\thelemma}{\Alph{appendixc}.\arabic{lemma}}
        \renewcommand{\thetheorem}{\Alph{appendixc}.\arabic{theorem}}
        \renewcommand{\thedefinition}{\Alph{appendixc}.\arabic{definition}}
        \renewcommand{\thecorollary}{\Alph{appendixc}.\arabic{corollary}}
        \renewcommand{\theequation}{\Alph{appendixc}.\arabic{equation}}
        \noindent{\tenbf Appendix \theappendixc #1}\par\vspace{5pt}}
\newcommand{\subappendix}[1] {\vspace{12pt}
        \refstepcounter{subappendixc}
        \noindent{\bf Appendix \thesubappendixc. {\kern1pt \bfit #1}}
	\par\vspace{5pt}}
\newcommand{\subsubappendix}[1] {\vspace{12pt}
        \refstepcounter{subsubappendixc}
        \noindent{\rm Appendix \thesubsubappendixc. {\kern1pt \tenit #1}}
	\par\vspace{5pt}}

\newcommand{\textlineskip}{\baselineskip=13pt}
\newcommand{\smalllineskip}{\baselineskip=10pt}

\def\eightcirc{
\begin{picture}(0,0)
\put(4.4,1.8){\circle{6.5}}
\end{picture}}
\def\eightcopyright{\eightcirc\kern2.7pt\hbox{\eightrm c}}

\newcommand{\copyrightheading}[1]
	{\vspace*{-2.5cm}\smalllineskip{\flushleft
	{\footnotesize Modern Physics Letters A, #1}\\
	{\footnotesize $\eightcopyright$\, World Scientific Publishing
	 Company}\\
	 }}

\def\abstracts#1#2#3{{
	\centering{\begin{minipage}{4.5in}\footnotesize\baselineskip=10pt
	\parindent=0pt #1\par
	\parindent=15pt #2\par
	\parindent=15pt #3
	\end{minipage}}\par}}

\newcommand{\bibit}{\nineit}
\newcommand{\bibbf}{\ninebf}
\renewenvironment{thebibliography}[1]
	{\frenchspacing
	 \ninerm\baselineskip=11pt
	 \begin{list}{\arabic{enumi}.}
        {\usecounter{enumi}\setlength{\parsep}{0pt}
	 \setlength{\leftmargin 12.7pt}{\rightmargin 0pt} %FOR 1--9 ITEMS
         \setlength{\itemsep}{0pt} \settowidth
	{\labelwidth}{#1.}\sloppy}}{\end{list}}

\newcounter{itemlistc}
\newcounter{romanlistc}
\newcounter{alphlistc}
\newcounter{arabiclistc}

\newcommand{\fcaption}[1]{
        \refstepcounter{figure}
        \setbox\@tempboxa = \hbox{\footnotesize Fig.~\thefigure. #1}
        \ifdim \wd\@tempboxa > 5in
           {\begin{center}
        \parbox{5in}{\footnotesize\smalllineskip Fig.~\thefigure. #1}
            \end{center}}
        \else
             {\begin{center}
             {\footnotesize Fig.~\thefigure. #1}
              \end{center}}
        \fi}

\newcommand{\tcaption}[1]{
        \refstepcounter{table}
        \setbox\@tempboxa = \hbox{\footnotesize Table~\thetable. #1}
        \ifdim \wd\@tempboxa > 5in
           {\begin{center}
        \parbox{5in}{\footnotesize\smalllineskip Table~\thetable. #1}
            \end{center}}
        \else
             {\begin{center}
             {\footnotesize Table~\thetable. #1}
              \end{center}}
        \fi}

\def\@citex[#1]#2{\if@filesw\immediate\write\@auxout
	{\string\citation{#2}}\fi
\def\@citea{}\@cite{\@for\@citeb:=#2\do
	{\@citea\def\@citea{,}\@ifundefined
	{b@\@citeb}{{\bf ?}\@warning
	{Citation `\@citeb' on page \thepage \space undefined}}
	{\csname b@\@citeb\endcsname}}}{#1}}

\newif\if@cghi
\def\cite{\@cghitrue\@ifnextchar [{\@tempswatrue
	\@citex}{\@tempswafalse\@citex[]}}
\def\citelow{\@cghifalse\@ifnextchar [{\@tempswatrue
	\@citex}{\@tempswafalse\@citex[]}}
\def\@cite#1#2{{$\null^{#1}$\if@tempswa\typeout
	{IJCGA warning: optional citation argument
	ignored: `#2'} \fi}}

\def\pmb#1{\setbox0=\hbox{#1}
	\kern-.025em\copy0\kern-\wd0
	\kern.05em\copy0\kern-\wd0
	\kern-.025em\raise.0433em\box0}

\def\fnm#1{$^{\mbox{\scriptsize #1}}$}
\def\fnt#1#2{\footnotetext{\kern-.3em
	{$^{\mbox{\scriptsize #1}}$}{#2}}}

\font\tenrm=cmr10
\font\tenit=cmti10
\font\tenbf=cmbx10
\font\bfit=cmbxti10 at 10pt
\font\ninerm=cmr9
\font\nineit=cmti9
\font\ninebf=cmbx9
\font\eightrm=cmr8

\def\qed{\hbox{${\vcenter{\vbox{			%HOLLOW SQUARE
   \hrule height 0.4pt\hbox{\vrule width 0.4pt height 6pt
   \kern5pt\vrule width 0.4pt}\hrule height 0.4pt}}}$}}

\renewcommand{\thefootnote}{\fnsymbol{footnote}}	%USE SYMBOLIC FOOTNOTE

\def\@citexs[#1]#2{\if@filesw\immediate\write\@auxout
    {\string\citation{#2}}\fi
\def\@citeas{}\@cites{\@for\@citebs:=#2\do
    {\@citeas\def\@citeas{,}\@ifundefined
    {b@\@citebs}{{\bf ?}\@warning
    {Citation `\@citebs' on page \thepage \space undefined}}
    {\csname b@\@citebs\endcsname}}}{#1}}

\newif\if@cghi
\def\cites{\@cghitrue\@ifnextchar [{\@tempswatrue
    \@citexs}{\@tempswafalse\@citexs[]}}
\def\citelows{\@cghifalse\@ifnextchar [{\@tempswatrue
    \@citexs}{\@tempswafalse\@citex[]}}
\def\@cites#1#2{{$\null {#1}$\if@tempswa\typeout
    {IJCGA warning: optional citation argument
    ignored: `#2'} \fi}}

\newcommand{\ind}[2]{^{#1}_{\mbox{\scriptsize #2}}}
\newcommand{\n}{{^{\mbox{\tiny N}}}\!}
\newcommand{\al}[2]{\alpha\ind{#1}{#2}}
\newcommand{\tal}[2]{\widetilde{\alpha}\ind{#1}{#2}}
\newcommand{\bt}[2]{\beta\ind{#1}{#2}}
\newcommand{\tro}[2]{\widetilde{\rho}\ind{#1}{#2}}
\def\nf{n_{\mbox{\scriptsize f}}}
\def\LQCD{$\Lambda_{\mbox{\scriptsize QCD}}$ }
\def\MSbar{$\overline{\mbox{MS}}$ }

\begin{document}
\setlength{\textheight}{7.7truein}  %for 2nd page onwards
%\runninghead{}{}
\normalsize\textlineskip
\thispagestyle{empty}
\setcounter{page}{1}

\copyrightheading{}           %{Vol. 0, No.0 (1992) 000--000}

\vspace*{0.88truein}
%\fpage{1}
\centerline{\bf NEW ANALYTIC RUNNING COUPLING IN QCD:}
\baselineskip=13pt
\centerline{\bf HIGHER LOOP LEVELS}
\vspace*{0.37truein}
\centerline{\footnotesize A.\ V.\ NESTERENKO\footnote{
E-mail: nesterav@thsun1.jinr.ru}}
\baselineskip=12pt
\centerline{\footnotesize\it Department of Physics,
Moscow State University}
\baselineskip=10pt
\centerline{\footnotesize\it Vorobjovy Gory, Moscow, 119899, Russia}
\vspace*{10pt}

\centerline{\footnotesize I.\ L.\ SOLOVTSOV\footnote{
E-mail: solovtso@thsun1.jinr.ru}}
\baselineskip=12pt
\centerline{\footnotesize\it Bogoliubov Laboratory of
Theoretical Physics,}
\baselineskip=10pt
\centerline{\footnotesize\it Joint Institute for Nuclear Research}
\centerline{\footnotesize\it Dubna, 141980, Russia}
\vspace*{0.225truein}

%\publisher{(received date)}{(revised date)}

\vspace*{0.21truein}
\abstracts{The properties of the new analytic running coupling are
investigated at the higher loop levels. The expression for this
invariant charge, independent of the normalization point, is obtained
by invoking the asymptotic freedom condition. It is shown that at any
loop level the relevant $\beta$ function has the universal behaviors
at small and large values of the invariant charge. Due to this
feature the new analytic running coupling possesses the universal
asymptotics both in the ultraviolet and infrared regions irrespective
of the loop level. The consistency of the model considered with the
general definition of the QCD invariant charge is shown.}{}{}

%\vspace*{10pt}
%\keywords{}

\setcounter{footnote}{0}
\renewcommand{\thefootnote}{\alph{footnote}}

\vspace*{1pt}\textlineskip
\section{Introduction}
     The description of hadron interaction in the infrared region
remains an actual problem of Quantum Chromodynamics (QCD). Since the
standard perturbation theory provides no definite answer on this
problem, a variety of nonperturbative methods is usually invoked for
the comprehensive investigation of this matter. The point is that the
renormalization group (RG) summation leads to the violation of the
proper analytic properties of the relevant physical quantities, that
contradicts the general principles of the theory. For instance, the
ghost pole appears in the expression for the running coupling at the
one-loop level. Therefore, in order to improve this situation one has
to involve into consideration the condition of analyticity.

     Being based on the first principles of the local Quantum Field
Theory, the analytic approach seeks to recover the violated after RG
summation proper analytic properties of the relevant physical
quantities. Its original ideas were formulated\cite{Redm,BLS} in the
framework of Quantum Electrodynamics (QED) in the late 1950's. The
basic idea of this approach is an explicit imposition of the
causality condition which implies the requirement of the analyticity
in the $q^2$ variable for the relevant physical quantities. The
analytic approach has recently been extended to QCD\cite{ShSol} and
applied to the `analytization' of the perturbative series for the QCD
observables.\cite{SolSh1} The term `analytization' means the
recovering of the proper analytic properties in the $q^2$ variable by
making use of the K\"all\'en--Lehmann spectral representation
\begin{equation}
\Bigl\{\mbox{\sf A}(q^2)\Bigr\}_{\mbox{$\!${\small an}}} \equiv
\int_{0}^{\infty}\! \frac{\varrho(\sigma)}{\sigma+q^2}\, d \sigma
\end{equation}
with the spectral density $\varrho(\sigma)$ determined by the initial
(perturbative) expression for a quantity {\sf A}$(q^2)$:
\begin{equation}
\varrho(\sigma) \equiv \frac{1}{2 \pi i} \lim_{\varepsilon \to 0_{+}}
\Bigl[\mbox{\sf A}(-\sigma-i \varepsilon) -
\mbox{\sf A}(-\sigma+i \varepsilon)\Bigr], \quad \sigma \ge 0.
\end{equation}

     The analytic approach has recently been applied to the
`improvement' of the $\beta$ function perturbative
expansion.\cite{PRD,NPQCD01} In accordance with the model proposed in
Refs.~\cites{PRD,NARCSTR}, the new analytic running coupling (NARC)
is the solution of the renormalization group equation for the
invariant charge with the analytized $\beta$ function. This running
coupling possesses a number of appealing features.\cite{NPQCD01,MPLA}
Its most important advantages are the following. First of all, the
new analytic running coupling has no unphysical singularities.
Further, the NARC explicitly involves the ultraviolet (UV) asymptotic
freedom with the infrared (IR) enhancement in a single expression. It
is worth noting here that there is a number of evidences for such a
behavior of the QCD invariant charge. In particular, the recent
lattice simulations\cite{UKQCD,ALPHA} as well as the solution of the
Schwinger--Dyson equations\cite{AlekArbu} testify to the IR
enhancement of the QCD running coupling. Furthermore, it was
demonstrated recently\cite{PRD,ConfIV} that the NARC provides the
quark confinement, the one-gluon exchange model being employed.
Remarkably, the new model for the analytic running coupling has no
additional parameters, i.e., similarly to the perturbative approach,
\LQCD remains the basic characterizing parameter of the theory.

     In this letter the investigation\cite{MPLA} of the new analytic
running coupling is continued. The primary objective is to study the
properties of the NARC at the higher loop levels. Since we have only
the integral representation for the NARC here, its straightforward
investigation turns out to be rather complicated. Nevertheless, by
examining the properties of the $\beta$ function corresponding to the
NARC one succeeded in the description of the asymptotic behavior of
the new analytic running coupling at the higher loop levels.

     The layout of the letter is as follows. In Sec.~2 the new
analytic running coupling is briefly discussed. By invoking the
asymptotic freedom condition the expression for the NARC, independent
of the normalization point, is obtained. Further, the conclusions
concerning the higher loop and scheme stability of the current
approach are drawn. In Sec.~3 the $\beta$ function corresponding to
the new analytic running coupling at the higher loop levels is
investigated. It is shown that this $\beta$ function has universal
behaviors both at the small and large values of the invariant charge,
irrespective of the loop level considered. This results in the
universal asymptotics of the new analytic running coupling both in
the UV and IR regions at any loop level. In Sec.~4 the compatibility
of the new model for the analytic running coupling with the general
definition of the QCD invariant charge is shown. The ways of the
analyticity requirement implementation are discussed. In Sec.~5 the
obtained results are summarized.

\section{New Analytic Running Coupling in QCD}
     The complementation of the $\beta$ function perturbative
expansion with the analyticity requirement leads to the analytized RG
equation for the new analytic running coupling (see
Refs.~\cites{PRD,NARCSTR} for the details). At the $\ell$-loop level
this equation acquires the form
\begin{equation}
\label{RGEqnNARC}
\frac{d \,\ln\bigl[\n\tal{(\ell)}{an}(\mu^2)\bigr]}{d \,\ln \mu^2} =
- \left\{
\sum_{j=0}^{\ell-1} B_j \Bigl[\tal{(\ell)}{s}(\mu^2)\Bigr]^{j+1}
\right\}_{\mbox{$\!\!$\small an}},
\quad B_j=\frac{\beta_{j}}{\beta_{0}^{j+1}}.
\end{equation}
Here $\n\al{(\ell)}{an}(\mu^2)$ is the new analytic running coupling
at the $\ell$-loop level, $\al{(\ell)}{s}(\mu^2)$ denotes the
$\ell$-loop perturbative running coupling, $\widetilde{\alpha}(\mu^2)
= \alpha(\mu^2)\, \beta_{0}/(4\pi)$, $\beta_{0} = 11 - 2\,\nf\,/\,3$
and $\beta_{1}=102 - 38 \,\nf\,/\,3$ are the standard coefficients
for the $\beta$ function expansion, and $\nf$ is the number of active
quarks. At the one-loop level Eq.\ (\ref{RGEqnNARC}) can be
integrated explicitly with the result\cite{PRD}
\begin{equation}
\label{NARC1L}
\n\al{(1)}{an}(q^2) = \frac{4\pi}{\beta_0}\,\frac{z-1}{z \,\ln z},
\qquad z=\frac{q^2}{\Lambda^2}.
\end{equation}
The properties of the one-loop NARC (\ref{NARC1L}) have been studied
in details in Ref.~\cites{MPLA}. It is worth to emphasize here that
the new analytic running coupling (\ref{NARC1L}) possesses the
following significant feature
\begin{equation}
\n\al{(1)}{an}(q^2) = \frac{\Lambda^2}{q^2}\,
\n\al{(1)}{an}\!\left(\frac{\Lambda^4}{q^2}\right).
\end{equation}
Recently it has been shown\cite{Schrempp} that this symmetry of the
QCD invariant charge precisely corresponds to the conformal inversion
symmetry of the instanton size distribution.\cite{UKQCD} In
particular, this implies that one can also derive the expression
(\ref{NARC1L}) for the QCD running coupling {\em proceeding from the
entirely different motivations}. It is interesting to note also
that the relation
\begin{equation}
\frac{\n\bt{(1)}{an}\!\left(\n\al{(1)}{an}(q^2)\right)}
{\n\tal{(1)}{an}(q^2)} +
\frac{\n\bt{(1)}{an}\!\left(\n\al{(1)}{an}(\Lambda^4/q^2)\right)}
{\n\tal{(1)}{an}(\Lambda^4/q^2)} = - 1
\end{equation}
holds for all $q^2 > 0$ in the framework of the model proposed.
Here $\n\al{(1)}{an}(q^2)$ is the invariant charge (\ref{NARC1L})
and $\n\bt{(1)}{an}$ denotes the relevant $\beta$ function
(see Eq.~(\ref{NBeta1L}) further).

     Let us turn to the higher loop levels. From the very beginning
one should note that the solution of Eq.~(\ref{RGEqnNARC}) is
determined up to a constant factor due to the logarithmic derivative
on its left-hand side. In previous studies\fnm{a}\fnt{a}{ In
Ref.~\cites{MPLA} the factor $z_0/z$ has been taken out from the
exponent of Eq.~(\ref{NARCHLNorm}) due to the property
$\int_{0}^{\infty}{\cal R}^{(\ell)}(\sigma)\,\sigma^{-1}\, d\sigma =
1$ for all $\ell$.} this problem has been eliminated by normalization
of the solution of Eq.~(\ref{RGEqnNARC}) to its value at a
point~$q_0^2$:
\begin{equation}
\label{NARCHLNorm}
\n\tal{(\ell)}{an}(q^2) = \n\tal{(\ell)}{an}(q_0^2)\,
\exp\!\left[\int_{0}^{\infty}\!
{\cal R}^{(\ell)}(\sigma)
\,\ln\!\left(\frac{1+\sigma/z}{1+\sigma/z_0}\right)
\frac{d \sigma}{\sigma}\right],\quad
z_0=\frac{q_0^2}{\Lambda^2},
\end{equation}
where
\begin{equation}
\label{RDef}
{\cal R}^{(\ell)}(\sigma) = \frac{1}{2 \pi i}
\lim_{\varepsilon \to 0_{+}}
\sum_{j=0}^{\ell-1} B_j
\biggl\{\Bigl[\tal{(\ell)}{s}(-\sigma-i \varepsilon)\Bigr]^{j+1}
-\Bigl[\tal{(\ell)}{s}(-\sigma+i \varepsilon)\Bigr]^{j+1}\biggr\}.
\end{equation}
However, in some cases it turns out to be more convenient to deal
with the explicit expression for the running coupling independent of
the normalization point.

     In this letter we propose a simple physical method for removing
the ambiguity mentioned above. Indeed, this can easily be achieved by
involving the condition of the asymptotic freedom,\fnm{b}\fnt{b}{ In
particular, this was used in Ref.~\cites{NPQCD01} when evaluating the
normalization coefficients.} namely $\n\tal{(\ell)}{an}(q^2)
\to \tal{(\ell)}{s}(q^2)$ when $q^2 \to \infty$ (in fact, this has
already been used in Eq.~(\ref{NARC1L})). It is worth noting that
this method does not violate the renormalization invariance of
Eq.~(\ref{NARCHLNorm}). In general, one becomes able to derive the
integral representation for the $\ell$-loop new analytic running
coupling, independent of the normalization point, by invoking the
similar condition $\n\tal{(\ell)}{an}(q^2) \to \n\tal{(1)}{an}(q^2)$
when $q^2 \to \infty$. For this purpose let us consider the ratio of
the $\ell$-loop new analytic running coupling (\ref{NARCHLNorm})
normalized at a point $q_0^2$ to the one-loop NARC written in the
form (\ref{NARCHLNorm}) and normalized at the same point $q_0^2$:
\begin{equation}
\frac{\n\tal{(\ell)}{an}(q^2)}{\n\tal{(1)}{an}(q^2)} =
\frac{\n\tal{(\ell)}{an}(q_0^2)}{\n\tal{(1)}{an}(q_0^2)}\,
\exp\!\left[
\int_{0}^{\infty}\!\! \Delta {\cal R}^{(\ell)}(\sigma)\,
\ln\!\left(\frac{1 + \sigma/z}{1 + \sigma/z_0}\right)
\frac{d \sigma}{\sigma}\right],
\end{equation}
where $\Delta {\cal R}^{(\ell)}(\sigma) = {\cal R}^{(\ell)}(\sigma) -
{\cal R}^{(1)}(\sigma)$. Proceeding to the limit $q_0^2 \to \infty$,
we arrive at the expression for the $\ell$-loop new analytic running
coupling
\begin{equation}
\label{NARCHighLoop}
\n\al{(\ell)}{an}(q^2) = \frac{4\pi}{\beta_0}\,\frac{z-1}{z \, \ln z}\,
\exp\left[\int_{0}^{\infty}\!\!
\Delta {\cal R}^{(\ell)}(\sigma)\,
\ln\!\left(1 + \frac{\sigma}{z}\right) \frac{d \sigma}{\sigma}
\right].
\end{equation}
It is worth noting that there is the integral representation of the
K\"all\'en--Lehmann type for the new analytic running coupling
\begin{equation}
\n\tal{(\ell)}{an}(q^2) = \int_{0}^{\infty}
\frac{\n\tro{(\ell)}{}(\sigma)}{\sigma + z}\, d \sigma,
\end{equation}
where
\begin{eqnarray}
\label{NRhoDef}
\n\tro{(\ell)}{}(\sigma) &=& \n\tro{(1)}{}(\sigma) \,
\exp\!\left[ \int_{0}^{\infty}\! \Delta {\cal R}^{(\ell)}(\zeta) \,
\ln \left| 1 - \frac{\zeta}{\sigma}\right| \,
\frac{d \zeta}{\zeta} \right] \nonumber \\
&& \times \left[\cos \psi^{(\ell)}(\sigma) +
\frac{\ln \sigma}{\pi} \sin \psi^{(\ell)}(\sigma) \right]
\end{eqnarray}
is the $\ell$-loop spectral density,
\begin{equation}
\n\tro{(1)}{}(\sigma) = \left( 1 + \frac{1}{\sigma} \right) \,
\frac{1}{\ln^2 \sigma + \pi^2}
\end{equation}
is the one-loop spectral density, and
\begin{equation}
\psi^{(\ell)}(\sigma) = \pi \int_{\sigma}^{\infty}\!
\Delta {\cal R}^{(\ell)}(\zeta) \, \frac{d \zeta}{\zeta}.
\end{equation}
In the exponent of Eq.~(\ref{NRhoDef}) the principle value of the
integral is assumed. However, in the practical use the expression
(\ref{NARCHighLoop}) turns out to be more convenient (at least,
beyond the one-loop level).

     One of the important features of the new analytic running
coupling is the absence of unphysical singularities at any loop
level.  Furthermore, it involves both the asymptotic freedom behavior
and the IR enhancement in a single expression. This feature enables
one to describe a broad range of physical phenomena including both
the standard perturbative and the intrinsically nonperturbative ones
(see Ref.~\cites{NPQCD01} for the details). In particular, it has
been shown\cite{PRD,ConfIV} that in the framework of the one-gluon
exchange model the new analytic running coupling (\ref{NARC1L})
explicitly leads to the rising at large distances static
quark-antiquark potential. Let us emphasize here that the additional
parameters are not introduced in the theory. The detailed description
of the properties and advantages of the NARC (\ref{NARC1L}) is given
in the papers.\cite{NPQCD01,MPLA}

\begin{figure}[ht]
\noindent
\vspace*{13pt}
\centerline{\epsfig{file=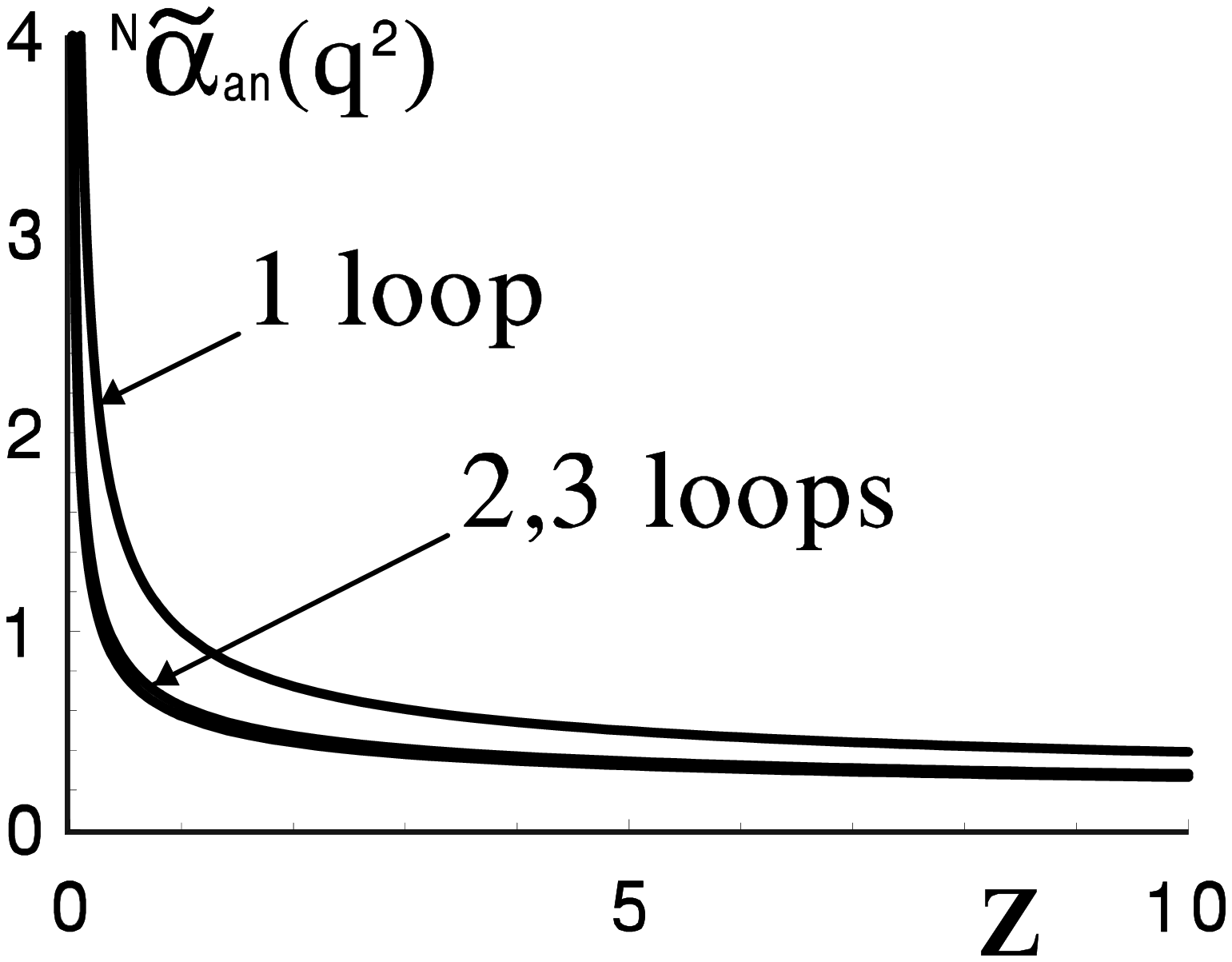, width=75mm}}
\vspace*{13pt}
\fcaption{The new analytic running coupling at different loop levels,
$z=q^2/\Lambda^2$.}
\label{Fig:NARC}
\end{figure}

     The Figure \ref{Fig:NARC} shows the new analytic running
coupling (\ref{NARCHighLoop}) at the one-, two- and three-loop
levels. It is obvious that NARC possesses the higher
loop stability. Thus, the curves corresponding to the two-, and
three-loop levels are practically indistinguishable. Proceeding from
this one can also draw the conclusion concerning the scheme stability
of the current approach. In particular, in Fig.~\ref{Fig:NARC} the
curve corresponding to the three-loop new analytic running coupling
$\n\tal{(3)}{an}$ is plotted by making use of the coefficient
$\beta_2 = 2857/2 - 5033\, \nf/18 + 325\, \nf^2/54$ computed\cite{TVZ}
in the
\MSbar scheme. The account of the third term on the right-hand side
of Eq.~(\ref{RGEqnNARC}) does not lead to a valuable quantitative
variation of its solution in comparison with the two-loop
approximation. Therefore it is clear that using the coefficient
$\beta_2$ computed in another subtraction scheme\fnm{c}\fnt{c}{ At
least, in schemes that do not have unnaturally large expansion
coefficients (see Ref.~\cites{Rac} and references therein for the
detailed discussion of this matter).} does not lead to significant
variation of the solution to Eq.~(\ref{RGEqnNARC}) in comparison with
the considered case of the \MSbar scheme. This statement follows also
from the fact that the contribution of every subsequent term on the
right-hand side of Eq.~(\ref{RGEqnNARC}) is substantially suppressed
by the contributions of the preceding ones.

     Since we have only the integral representation for the new
analytic running coupling at the higher loop levels, its
straightforward investigation becomes rather complicated.
Nevertheless, the examining of the $\beta$ function corresponding to
the NARC enables one to elucidate a number of important questions, in
particular, the asymptotic behavior of the new analytic running
coupling.

\section{The $\beta$ Function: Higher Loop Levels}
\begin{figure}[ht]
\noindent
\vspace*{13pt}
\centerline{\epsfig{file=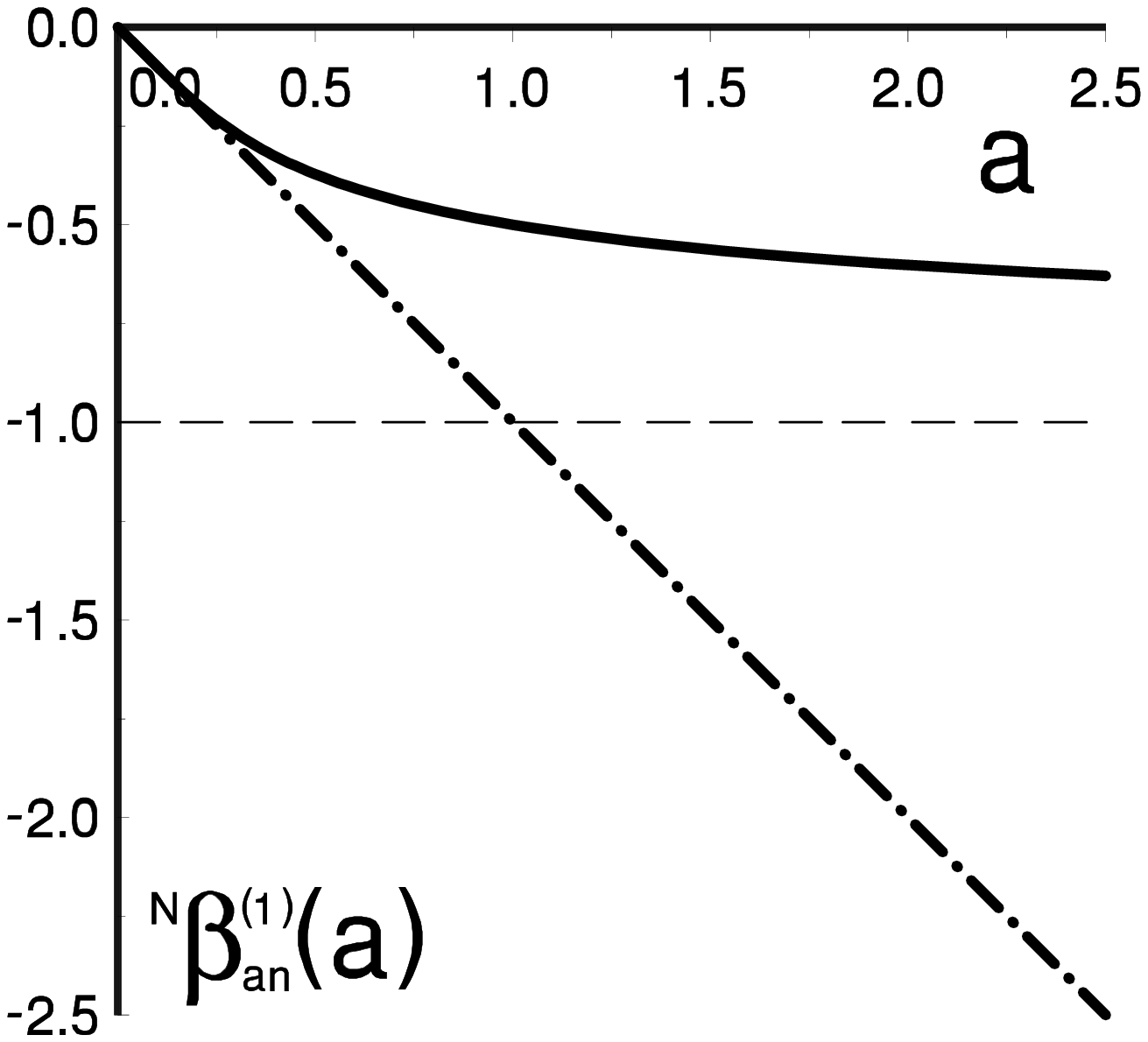, width=55mm}}
\vspace*{13pt}
\fcaption{The $\beta$ function corresponding to the one-loop new
analytic running coupling (solid curve). The relevant perturbative
result is shown by the dot-dashed line.}
\label{Fig:NABeta1L}
\end{figure}

     In the previous letter\cite{MPLA} the $\beta$ function
\begin{equation}
\label{BetaGenDef}
\beta(\alpha) = \frac{d \,\ln\alpha(\mu^2)}{d \,\ln\mu^2}
\end{equation}
corresponding to the one-loop new analytic running coupling
(\ref{NARC1L}) has been derived\fnm{d}\fnt{d}{ In Ref.~\cites{MPLA} the
definition $\beta(\alpha) = d\,\alpha(\mu^2)/d\,\ln\mu^2$ was used.}
and studied in details. Due to the explicit expression for the NARC
at the one-loop level, one succeeded in performing the relevant
investigation manifestly. Thus, the $\beta$ function
(\ref{BetaGenDef}) corresponding to the one-loop NARC (\ref{NARC1L})
acquires the form
\begin{equation}
\label{NBeta1L}
\n\bt{(1)}{an}(a) =
\frac{1-N(a)/a}{\ln\left[N(a)\right]},\quad
a(\mu^2) \equiv \n\tal{(1)}{an}(\mu^2),
\end{equation}
where the function $N(a)$
\begin{equation}
\label{NDef}
N(a) = \cases{
N_{0}(a),  & $0 < a \le 1$, \cr
N_{-1}(a), & $1 < a$, \cr}
\quad
N_{k}(a) = -a \, W_{k}\!
\left[-\frac{1}{a}\,\exp\!\left(-\frac{1}{a}\right)\right]
\end{equation}
is defined in terms of the many--valued Lambert $W$
function\fnm{e}\fnt{e}{ In definitions (\ref{NDef}) and (\ref{WEqDef})
$k$ denotes the branch index of the Lambert $W$ function.}
\begin{equation}
\label{WEqDef}
W_{k}(x)\,\exp\left[W_{k}(x)\right]=x
\end{equation}
(see Ref.~\cites{MPLA} for the details). Figure \ref{Fig:NABeta1L}
presents the $\beta$ function (\ref{NBeta1L}) and the result
corresponding to the one-loop perturbative running coupling
$\tal{(1)}{s}(q^2) = 1/\ln z$, namely $\bt{(1)}{s}(a) = -a$.

     The investigation\cite{MPLA} of the properties of the function
$N(a)$ defined in Eq.~(\ref{NDef}) enables one to find the explicit
asymptotic behavior of the $\beta$ function (\ref{NBeta1L}). Thus,
for small values of the running coupling $\n\tal{(1)}{an}$
Eq.~(\ref{NBeta1L}) coincides with the well-known perturbative result
\begin{equation}
\label{NABetaSerOr}
\n\bt{(1)}{an}(a) = -a +
O \!\left[a^{-1}\exp\!\left(-\frac{1}{a}\right)\right],\quad
a \to 0_{+}.
\end{equation}
It is worth noting here that the second term in
Eq.~(\ref{NABetaSerOr}) points to the intrinsically nonperturbative
nature of the $\beta$ function (\ref{NBeta1L}). For large values of
$a$ one has
\begin{equation}
\label{NABetaSerInf}
\n\bt{(1)}{an}(a) = -1 +
O\!\left[\frac{\ln(\ln a)}{\ln^2 a}\right],\quad
a \to \infty.
\end{equation}
Such a behavior of the $\beta$ function provides the IR enhancement
of the invariant charge, namely (up to a logarithmic factor)
$\alpha(q^2) \sim 1/z$, when $z \to 0$. In turn, this leads in a
straightforward way to the confining static quark--antiquark
potential (see Refs.~\cites{PRD,ConfIV} for the details). Therefore,
at the one-loop level the $\beta$ function (\ref{NBeta1L}) explicitly
incorporates the UV asymptotic freedom and the IR enhancement of the
running coupling in a single expression.

\begin{figure}[th]
\noindent
\vspace*{13pt}
\centerline{\epsfig{file=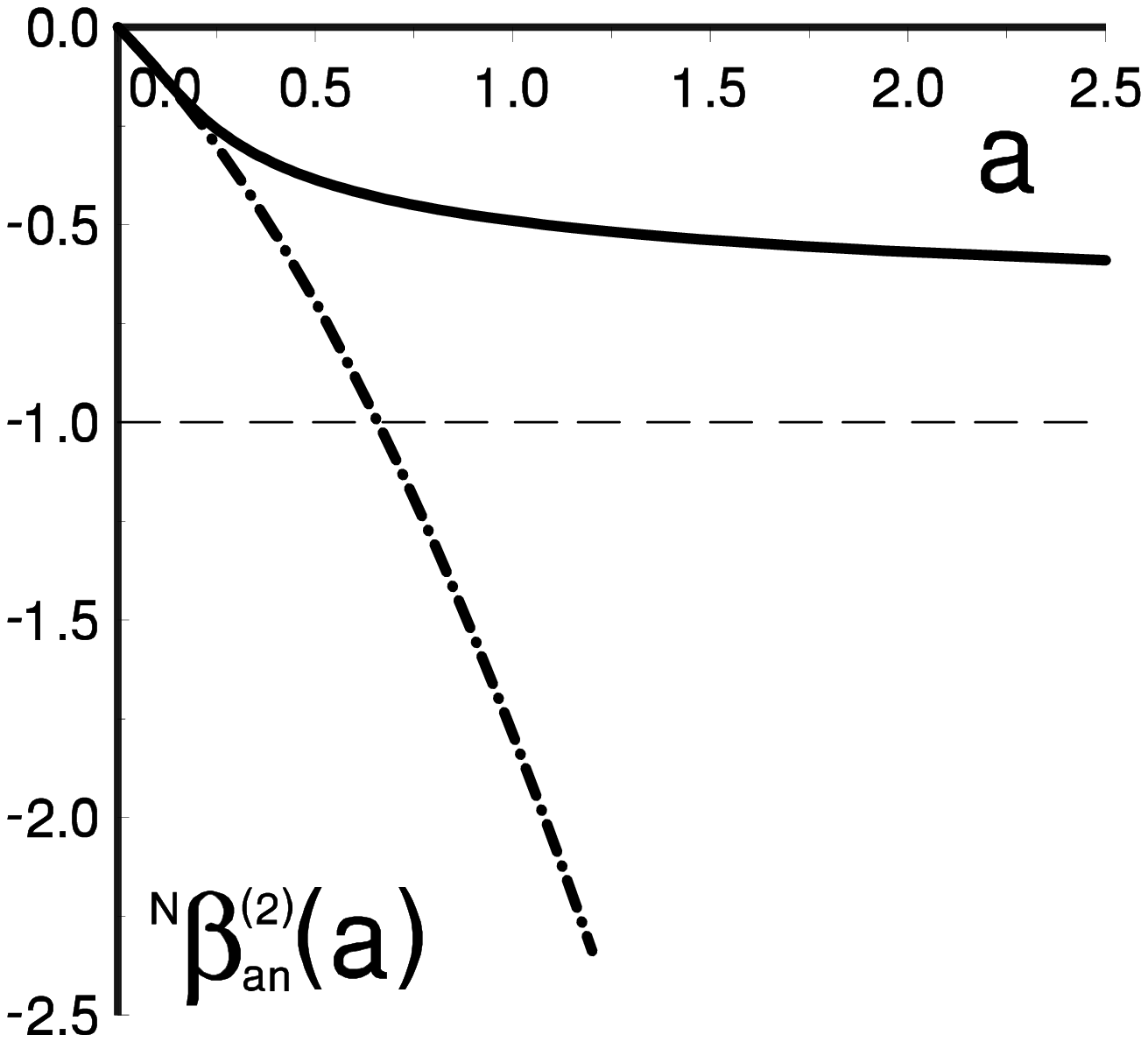, width=55mm}}
\vspace*{13pt}
\fcaption{The $\beta$ function corresponding to the two-loop new
analytic running coupling (solid curve). The relevant perturbative
result is shown as the dot-dashed curve.}
\label{Fig:NABeta2L}
\end{figure}

\begin{figure}[ht]
\noindent
\vspace*{13pt}
\centerline{\epsfig{file=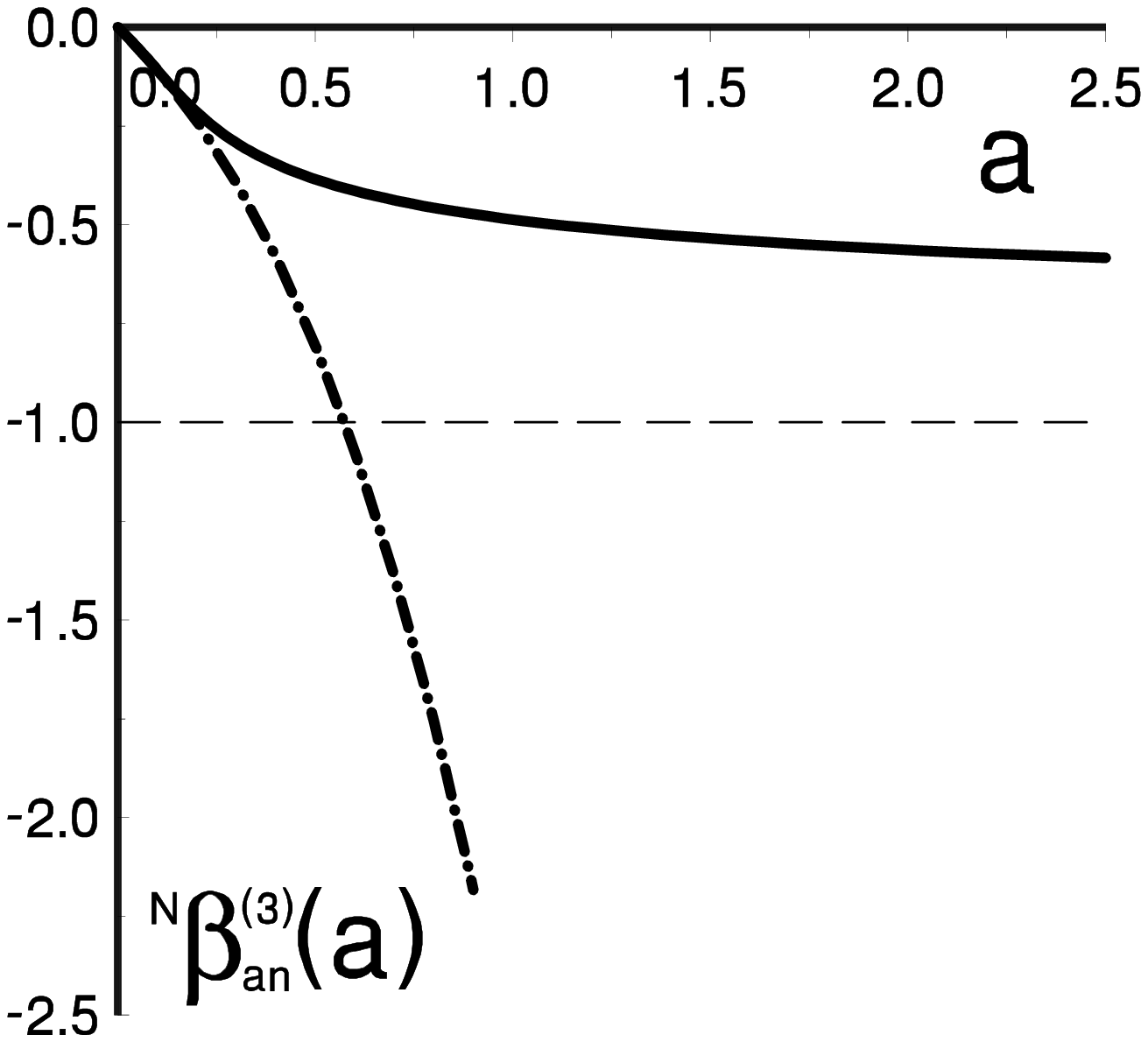, width=55mm}}
\vspace*{13pt}
\fcaption{The $\beta$ function corresponding to the three-loop new
analytic running coupling (solid curve). The relevant perturbative
result is shown as the dot-dashed curve.}
\label{Fig:NABeta3L}
\end{figure}

     Let us proceed to the higher loop levels. As it has been
mentioned in the previous section, here we have only the integral
representation for the new analytic running coupling
(\ref{NARCHighLoop}). This fact significantly complicates the
investigation and leads to the necessity of applying the numerical
methods. It is worth to mention here again that we originate in the
standard perturbative expansion for the $\beta$ function and
complement it with the analyticity requirement.  The results of the
computations corresponding to the two-, and three-loop levels are
shown in Figs.~\ref{Fig:NABeta2L} and \ref{Fig:NABeta3L},
respectively. Figure \ref{Fig:NABeta2L} presents the $\beta$
function $\n\bt{(2)}{an}(a)$ corresponding to the two-loop new
analytic running coupling $\n\tal{(2)}{an}(q^2)$ together with the
respective perturbative result $\bt{(2)}{s}(a) = -a - B_1 a^2$.
Figure \ref{Fig:NABeta3L} shows the analogous functions at the
three-loop level, namely $\n\bt{(3)}{an}(a)$ and $\bt{(3)}{s}(a) = -a
-B_1 a^2 -B_2 a^3$. It is clear from the figures \ref{Fig:NABeta1L},
\ref{Fig:NABeta2L}, and \ref{Fig:NABeta3L} that the $\beta$ function
corresponding to the new analytic running coupling coincides with its
perturbative analog in the region of small values of the invariant
charge at any loop level. In other words, in the framework of the
model in hand the complete recovering of the perturbative limit in
the UV region takes place.

\begin{figure}[t]
\noindent
\vspace*{13pt}
\centerline{\epsfig{file=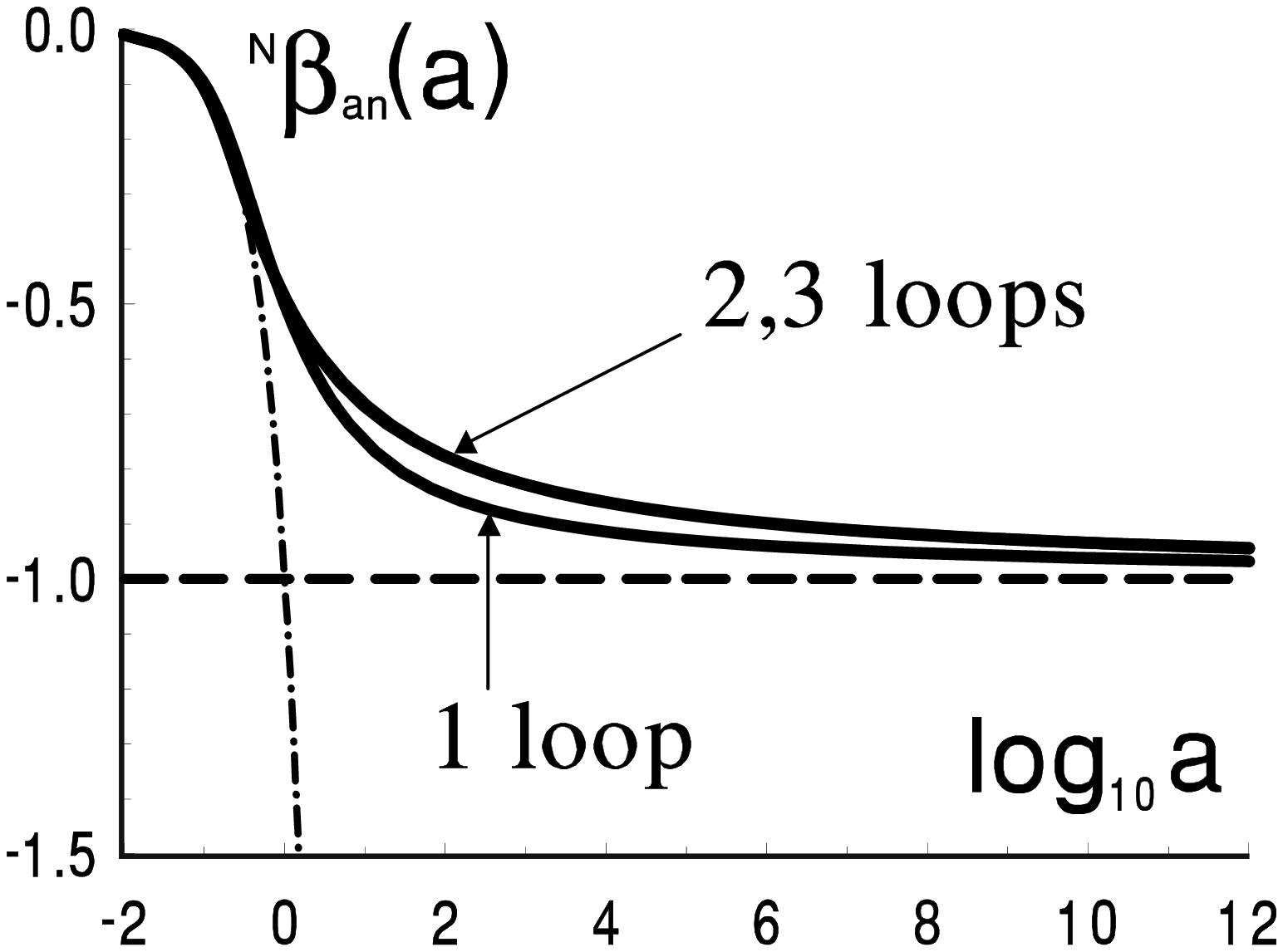, width=75mm}}
\vspace*{13pt}
\fcaption{The $\beta$ function corresponding to the new analytic
running coupling at different loop levels (solid curves). The
one-loop perturbative result is shown as the dot-dashed curve.}
\label{Fig:NABetaHL}
\end{figure}

     Let us turn now to the asymptotics of the $\beta$ function
$\n\bt{(\ell)}{an}(a)$ at large values of the running coupling.
Figure \ref{Fig:NABetaHL} presents the $\beta$ functions
$\n\bt{(\ell)}{an}(a),\,$ $\ell = 1,2,3$ and the one-loop
perturbative result $\bt{(1)}{s}(a) = -a$. This figure clearly
shows the perturbative limit at small $a$, as well as the universal
asymptotic behavior of the $\beta$ function corresponding to the NARC
$\n\bt{(\ell)}{an}(a) \to -1$, when $a \to \infty$.

     The latter statement can also be proved in an independent way.
Indeed, due to the IR enhancement of the new analytic running
coupling, the value of the right-hand side of Eq.~(\ref{RGEqnNARC})
(it is nothing but the $\beta$ function at the relevant loop level)
when $\mu^2 \to 0$ corresponds to the limit $\n\tal{(\ell)}{an} \to
\infty$. One can show that $\lim_{\,q^2 \to 0}
\left\{\tal{(\ell)}{s}(q^2)\right\} _{\!\mbox{\scriptsize an}} = 1$,
and $\lim_{\,q^2 \to 0} \Bigl\{\left[
\tal{(\ell)}{s}(q^2)\right]^{j+1}\Bigr\} _{\!\mbox{\scriptsize an}} =
0$ ($j \ge 1$ is a positive integer number), irrespective of the loop
level. Therefore, we infer that at any loop level the $\beta$
function corresponding to the new analytic running coupling tends to
the universal limit
\begin{equation}
\lim_{a \to \infty}\, \n\bt{(\ell)}{an}(a) = -1.
\end{equation}
As it has been mentioned above, such a behavior of the $\beta$
function leads to the IR enhancement of the invariant charge, namely
$\alpha(q^2) \sim 1/z$, when $z \to 0$.

     Thus, the $\beta$ function corresponding to the new analytic
running coupling has the universal asymptotic behaviors both at the
small ($\n\bt{(\ell)}{an}(a) \simeq -a$) and large
($\n\bt{(\ell)}{an}(a) \simeq -1$) values of the invariant charge
irrespective of the loop level. Therefore, the new analytic running
coupling possesses the universal asymptotics both in the UV
($\n\tal{(\ell)}{an}(q^2) \sim 1/\ln z$) and IR
($\n\tal{(\ell)}{an}(q^2) \sim 1/z$) regions at any loop level. In
particular, this implies that using the new analytic running coupling
(\ref{NARCHighLoop}) at the higher loop levels will also results in
the confining quark-antiquark potential.

\newpage
\section{Discussion}
     In general, the QCD invariant charge is defined as the product
of the corresponding Green functions and vertexes. For example, in
the transverse gauge the following definition
\begin{equation}
\label{InvChrgDef}
\alpha(q^2) = \alpha(\mu^2)\, G(q^2)\, g^2(q^2)
\end{equation}
takes place, where $\alpha(q^2)$ is the QCD running coupling,
$\alpha(\mu^2)$ denotes its value at a normalization point $\mu^2$,
$G(q^2)$ and $g(q^2)$ are dimensionless gluon and ghost propagators,
respectively. It is essential to note here that the Green functions
possess the proper analytic properties in the $q^2$ variable (namely,
there is the only left cut $q^2 \le 0$) before the RG summation.
Indeed, for the considering in this letter massless case a Green
function can be represented as the perturbative power series in
$\alpha(\mu^2)$ with the coefficients being a polynomial in $\ln
(q^2/\mu^2)$. Obviously, in this case the integral representation of
the K\"all\'en--Lehmann type must hold for any finite order of the
perturbative expansion for the Green function.\fnm{f}\fnt{f}{ One
should note that,
unlike the case of the QED photon propagator, the relevant spectral
function here can't be directly identified with the spectral density.}
The RG summation leads to violation of such a properties of
the Green functions (e.g., in the simplest one-loop case the ghost
pole appears). As it has already been mentioned above,
the analytic
approach to QCD seeks to recover the proper analytic properties of
the relevant quantities. However, for the consistency of
involving the analyticity condition with the definition of the
invariant charge (\ref{InvChrgDef}), one has to apply the
analytization procedure to the logarithm of the Green function.
Really, if the spectral function of the K\"all\'en--Lehmann
representation for the Green function $G$ is of a fixed sign (that is
true for the leading orders of perturbation theory), then $G$ has no
nulls in the complex $q^2$ plane. Hence, its logarithm, $\ln\,G$, can
also be represented as the spectral integral of the
K\"all\'en--Lehmann type.

     In the framework of perturbative approach the considered
above Green functions have the following form: $g(q^2) =
(1/\ln\,z)^{d_{g}}$ and $G(q^2) = (1/\ln\,z)^{d_{G}}$, where
$z=q^2/\Lambda^2$, $d_{g}$ and $d_{G}$ denote the corresponding
anomalous dimensions. For the latter the relationship $2 d_{g} +
d_{G} = 1$ holds, that plays the crucial role in the definition
(\ref{InvChrgDef}). Applying the analytization procedure to these
functions in the described above way, one arrives at the following
result: $g_{\mbox{\scriptsize an}}(q^2) = \left[(z-1)/(z\,\ln
z)\right] ^{d_{g}}$ and $G_{\mbox{\scriptsize an}}(q^2) =
\left[(z-1)/(z\,\ln z)\right] ^{d_{G}}$. Therefore, in this case the
invoking the analyticity condition does not affect the definition
of the invariant charge (\ref{InvChrgDef}). It is interesting to note
here that the similar situation takes place in the nonperturbative
$a$-expansion method\cite{VPT1,VPT2} also. Thus, we infer that the
NARC (\ref{NARC1L}) is consistent with the general definition of the
QCD invariant charge.

     It is worth to mention that there are different methods of removing
of the unphysical singularities from the running coupling in the
framework of the analytic approach to QCD. In original
work\cite{ShSol} the analytization procedure was applied to the
perturbative invariant charge in a straightforward way (see
Ref.~\cites{Sh} also). The model in hand\cite{PRD,NARCSTR} is based
on the complementation of the $\beta$ function perturbative expansion
with the analyticity condition. In both cases the additional
parameters are not introduced into the theory, the models being the
`minimal' ones in this sense. Of course, in the UV region these
models have identical behavior determined by the asymptotic freedom.
However, there is a qualitative distinction between them in the IR
region (see discussion in Ref.~\cites{NARCSTR} also).

     Let us note here that in equation (\ref{RGEqnNARC}) for the new
analytic running coupling the analytization of its right-hand side as
a whole is assumed. In turn, this results in the representation of
the corresponding $\beta$ function as a non-power series. As it has
already been shown,\cite{SolSh2,MSSY} it is this way of the
analyticity requirement involving that leads to the stability of
obtaining results in a broad range of energies. In particular, such
an implementation of the analytization procedure ultimately leads to
the universal IR behavior of the NARC.

     As it was mentioned above, the NARC incorporates the IR
enhancement with the UV asymptotic freedom in a single expression.
The recent results of the nonperturbative studies (namely, lattice
simulations\cite{UKQCD,ALPHA} and Schwinger--Dyson
equations\cite{AlekArbu}) testify to such a behavior of the QCD
invariant charge. For the completeness of the pattern let us also
note here that these nonperturbative methods, being the matter of
contemporary comprehensive investigations, provide no unique point of
view on the IR behavior of the QCD invariant charge.

\section{Conclusions}
     The properties of the new analytic running coupling are studied
at the higher loop levels. By making use of the asymptotic freedom
condition the expression for the NARC, independent of the
normalization point, is obtained. The conclusions about the loop and
scheme stability of the current approach are drawn. The $\beta$
function corresponding to the NARC is constructed and examined. It is
shown that the behaviors of this $\beta$ function at both its key
asymptotics (namely, when $a \to 0$ and when $a \to \infty$) are the
same at any loop level. This results, irrespective of the loop level,
in an universal asymptotics of the new analytic running coupling.
Namely, its UV behavior ($\n\tal{(\ell)}{an}(q^2) \sim 1/\ln z$, when
$q^2 \to \infty$) is determined by the asymptotic freedom, and there
is the IR enhancement of the new analytic running coupling
($\n\tal{(\ell)}{an}(q^2) \sim 1/z$, when $q^2 \to 0$). The
consistency of the model considered with the general definition of
the QCD invariant charge is shown.

\nonumsection{Acknowledgments}
     A.\ N.\ would like to thank Prof.\ F.\ Schrempp for the
stimulating comments. I.~S.\ is grateful to Profs.\ D.\ I.\ Kazakov
and D.\ V.\ Shirkov for interest in this work. The partial support of
RFBR (grants 99-01-00091, 99-02-17727, and 00-15-96691) is
appreciated.

\newpage
\nonumsection{References}


\begin{thebibliography}{000}
\bibitem{Redm} P.\ J.\ Redmond, {\bibit Phys.\ Rev.} {\bibbf 112},
         1404 (1958).
\bibitem{BLS} N.\ N.\ Bogoliubov, A.\ A.\ Logunov, and D.\ V.\
         Shirkov, {\bibit Zh.\ Eksp.\ Teor.\ Fiz.} {\bibbf 37}, 805
         (1959) [{\bibit Sov.\ Phys.\ JETP} {\bibbf 37}, 574 (1960)].
\bibitem{ShSol} D.\ V.\ Shirkov and I.\ L.\ Solovtsov, {\bibit Phys.\
         Rev.\ Lett.} {\bibbf 79}, 1209 (1997); hep-ph/9704333.
\bibitem{SolSh1} I.\ L.\ Solovtsov and D.\ V.\ Shirkov, {\bibit
         Theor.\ Math.\ Phys.} {\bibbf 120}, 1220 (1999);
         hep-ph/9909305.
\bibitem{PRD} A.\ V.\ Nesterenko, {\bibit Phys.\ Rev.} {\bibbf D62},
         094028 (2000); hep-ph/9912351.
\bibitem{NPQCD01} A.\ V.\ Nesterenko, in {\bibit Proceedings of the
         Sixth Workshop on Non-Perturbative Quantum Chromodynamics,
         Paris, France, 2001} (to be published); hep-ph/0106305.
\bibitem{NARCSTR} A.\ V.\ Nesterenko, {\bibit Phys.\ Rev.} {\bibbf D64},
         116009 (2001); hep-ph/0102124.
\bibitem{MPLA} A.\ V.\ Nesterenko, {\bibit Mod.\ Phys.\ Lett.}
         {\bibbf A15}, 2401 (2000); hep-ph/0102203.
\bibitem{UKQCD} UKQCD Collaboration, D.\ A.\ Smith and M.\ J.\ Teper,
         {\bibit Phys.\ Rev.} {\bibbf D58}, 014505 (1998);
         hep-lat/9801008.
\bibitem{ALPHA} ALPHA Collaboration, A.\ Bode {\bibit et al.},
         {\bibit Phys.\ Lett.} {\bibbf B515}, 49 (2001),
         hep-lat/0105003; J.\ Heitger {\bibit et al.},
         hep-lat/0110201.
\bibitem{AlekArbu} A.\ I.\ Alekseev and B.\ A.\ Arbuzov, {\bibit Mod.\
         Phys.\ Lett.} {\bibbf A13}, 1747 (1998); hep-ph/9704228.
\bibitem{ConfIV} A.\ V.\ Nesterenko, in {\bibit Proceedings of the
         4th International Conference on Quark Confinement and the
         Hadron Spectrum, Vienna, Austria, 2000} (in press);
         hep-ph/0010257.
\bibitem{Schrempp} F.\ Schrempp, hep-ph/0109032.
\bibitem{TVZ} O.\ V.\ Tarasov, A.\ A.\ Vladimirov and A.\ Yu.\ Zharkov,
         {\bibit Phys.\ Lett.} {\bibbf B93}, 429 (1980).
\bibitem{Rac} P.\ A.\ Raczka, {\bibit Z.\ Phys.} {\bibbf C65}, 481
         (1995); hep-ph/9407343.
\bibitem{VPT1} I.\ L.\ Solovtsov, {\bibit Phys.\ Lett.} {\bibbf B327},
         335 (1994); {\bibbf B340}, 245 (1994).
\bibitem{VPT2} A.\ N.\ Sisakyan and I.\ L.\ Solovtsov, {\bibit Phys.\
         Part.\ and Nucl.} {\bibbf 30}, 461 (1999).
\bibitem{Sh} D.\ V.\ Shirkov, {\bibit Eur.\ Phys.\ J.} {\bibbf C22},
         331 (2001); hep-ph/0107282.
\bibitem{SolSh2} I.\ L.\ Solovtsov and D.\ V.\ Shirkov, {\bibit Phys.\
         Lett.} {\bibbf B442}, 344 (1998); hep-ph/9711251.
\bibitem{MSSY} K.\ A.\ Milton, I.\ L.\ Solovtsov, O.\ P.\ Solovtsova,
         and V.\ I.\ Yasnov, {\bibit Eur.\ Phys.\ J.} {\bibbf C14},
         495 (2000); hep-ph/0003030.
\end{thebibliography}
\end{document}